\documentclass[prx,aps,twocolumn,superscriptaddress]{revtex4-2}
\usepackage[colorlinks=true,urlcolor=blue,citecolor=blue,linkcolor=blue]{hyperref}

\usepackage[T1]{fontenc}
\usepackage[latin9]{inputenc}
\usepackage{amssymb}
\usepackage{graphicx}
\usepackage{amsmath,color}
\usepackage{mathrsfs}
\usepackage{float}
\usepackage{indentfirst}
\usepackage{subfigure}
\usepackage{multirow}
\usepackage{tabu}
\usepackage{booktabs}
\usepackage{txfonts}
\usepackage{amsmath,amssymb,bbm}
\usepackage[normalem]{ulem}

\usepackage[euler]{textgreek}

\usepackage{graphicx}
\usepackage{bm}

\usepackage[dvipsnames]{xcolor}
\usepackage{soul}

\newcounter{nnnote}

{\newtheorem{note*}[nnnote]{Note}}

\newcommand{\Eq}[1]{Eq.~(\ref{#1})}
\newcommand{\<}{\langle}
\renewcommand{\>}{\rangle}
\newcommand{\id}{\mathbbm{1}}
\newcommand{\dtau}{\epsilon}
\newcommand{\tr}{\ensuremath{\operatorname{Tr}}}

\newcommand\redsout{\bgroup\markoverwith{\textcolor{red}{\rule[0.5ex]{2pt}{0.4pt}}}\ULon}
\newcommand{\dTRG}{$\partial$TRG }

\usepackage{scalerel}
\DeclareRobustCommand{\xjoinrel}{\mathrel{\mkern-3mu}}
\DeclareRobustCommand{\gong}{\hstretch{1.25} {\boldsymbol{\vdash \xjoinrel\dashv }}}
\DeclareRobustCommand{\wang}{\hstretch{1.25} {\boldsymbol{\vdash \xjoinrel \mathrel{+} \xjoinrel\dashv }}}
\DeclareRobustCommand{\+}{\hstretch{1.25} {\boldsymbol {\mathrel{+}}}}
\DeclareRobustCommand{\manyplus}{\hstretch{1.25} {\boldsymbol{ \mathrel{+}\xjoinrel \mathrel{+}\xjoinrel\mathrel{+} \xjoinrel\mathrel{+}}  }}
\DeclareRobustCommand{\l}{\hstretch{1.25} {\boldsymbol {\vdash }}}
\DeclareRobustCommand{\r}{\hstretch{1.25} {\boldsymbol {\dashv }}}
\DeclareRobustCommand{\tu}{\hstretch{1.25} {\boldsymbol {\mathrel{+} \xjoinrel\dashv }}}

\begin{document}

\preprint{APS/123-QED}

\title{Continuous Matrix Product Operator Approach to Finite Temperature Quantum States}%

\author{Wei Tang}
\affiliation{International Center for Quantum Materials, School of Physics, Peking University, Beijing 100871, China}

\author{Hong-Hao Tu}
\email{hong-hao.tu@tu-dresden.de}
\affiliation{Institute of Theoretical Physics, Technische Universit\"at Dresden, 01062 Dresden, Germany}

\author{Lei Wang}
\email{wanglei@iphy.ac.cn}
\affiliation{Beijing National Lab for Condensed Matter Physics and Institute of Physics, Chinese Academy of Sciences, Beijing 100190, China}
\affiliation{Songshan Lake Materials Laboratory, Dongguan, Guangdong 523808, China}

\date{\today}

\begin{abstract}
We present an algorithm for studying quantum systems at finite temperature using continuous matrix product operator representation.
The approach handles both short-range and long-range interactions in the thermodynamic limit without incurring any time discretization error.
Moreover, the approach provides direct access to physical observables including the specific heat, local susceptibility, and local spectral functions.
After verifying the method using the prototypical quantum $XXZ$ chains, we apply it to quantum Ising models with power-law decaying interactions and on the infinite cylinder, respectively. The approach offers predictions that are relevant to experiments in quantum simulators and the nuclear magnetic resonance spin-lattice relaxation rate.
\end{abstract}

\maketitle

{\em Introduction} --- Despite being used as a standard method for studying ground states of low-dimensional quantum systems~\cite{verstraete2008, schollwock2011, stoudenmire2012studying, orus2018}, tensor network approaches to thermal states are still under continuous development~\cite{Bursill1996, Wang1997, xiang1998thermodynamics, Shibata, PhysRevLett.93.207204, PhysRevLett.93.207205, Feiguin2005, White2009, Stoudenmire2010, PhysRevLett.106.127202, Xie2012, Czarnik2015, Chen2017i, Chen2018q, Kshetrimayum2019, Chung, Chen2019l, Chen2019m}.
Ideally, for translationally invariant systems we would like to have a method that directly works in the thermodynamical limit, handles long-range interactions and two-dimensional geometry well, has no imaginary-time discretization error, and even better, provides access to dynamical properties such as finite temperature spectral functions.

The existing approaches have made a trade-off in meeting either this or that from the wish list. The difficulties are that, although the partition function of a quantum system could be formally treated as a tensor network defined in a one-dimensional-higher space-time manifold with an imaginary-time direction, the tensor network has periodic boundary condition in the time direction with a periodicity given by the inverse temperature $\beta= 1/T$. Moreover, the tensor network is highly anisotropic as it is continuous along the imaginary-time direction. These features made it nontrivial to transferring those highly successful tensor network approaches developed for the ground-state calculation.

In this work, we present an approach to study quantum systems at finite temperature centered around the concept of continuous matrix product operators (cMPOs). The approach works in the thermodynamic limit and the continuous-time limit simultaneously. Besides the obvious advantages of eliminating the finite-size and time-discretization errors, the approach directly applies to systems with long-range interactions by virtue of the MPO representation. Moreover, a distinguishing feature of the present approach is that it offers ways to straightforwardly measure physical observables such as unequal-time correlation functions and dynamical responses at finite temperature. In analogy to the density matrix renormalization group (DMRG)~\cite{stoudenmire2012studying}, quasi-one-dimensional systems with a cylindrical geometry are also within the scope of the cMPO method, which opens a way to study  statistical and dynamical properties of two-dimensional quantum systems at finite temperature. 

The present approach stems from a compact MPO representation of the evolution operator for time evolving long-range interacting systems~\cite{Zaletel2015}. Despite being simple and elegant, the original MPO representation has an intrinsic time-discretization error, which was shown to be not accurate enough for practical calculations~\cite{Zaletel2015, Bruognolo2017a}. We show that one can eliminate the time discretization error by formally taking the continuous-time limit in the tensor network algorithm in the same spirit as the continuous-time quantum Monte Carlo (QMC) approaches~\cite{Beard1996,Prokofev1996}. A tensor network formed by cMPOs appears naturally in the continuous-time limit.  To contract such tensor network, one encounters the continuous matrix product state (cMPS)~\cite{Verstraete2010a} as the dominant eigenvector of the cMPO. Having such cMPS along the imaginary-time direction of finite temperature quantum systems has been anticipated in Refs.~\cite{Hastings2015, Tirrito2018}, but its implications for practical calculations have been largely unexplored.

{\em cMPO formulation} --- There is a general recipe to construct the MPO representation of a Hamiltonian $H$, including those with long-range interactions~\cite{McCulloch2007, Crosswhite2008a, Crosswhite2008, Frowis2010}. Building on such representation, one can also construct an accurate MPO representation for the evolution operator $e^{-\dtau H}$ provided the time step $\dtau$ is sufficiently small~\cite{Pirvu2010, Zaletel2015}. Since we will consider the limit of $\dtau \rightarrow 0$, we shall not be concerned about the time discretization error and write the evolution operator as a translationally invariant MPO, $e^{-\dtau H} = \cdots \manyplus \cdots$, where the vertical legs of the MPO tensor represent the $d$-dimensional physical Hilbert space at each site, and the horizontal legs carry $D$-dimensional virtual degrees of freedom. When viewing the left and right legs of the MPO tensor as matrix indices, the tensor at each site takes the same form~\cite{Zaletel2015}
\begin{equation}
{\+}_{ij} = \left(\begin{array}{cc} \id + \dtau {Q} & \sqrt{\dtau} \bm{L} \\ \sqrt{\dtau} \bm{R} &  \bm{P} \end{array}\right)_{ij}, \label{eq:cMPO}
\end{equation}
where the subscripts $i,j \in [1, D]$ are virtual indices and each matrix entry ${\+}_{ij}$ is an operator acting on the $d$-dimensional physical Hilbert space. In terms of the virtual indices, the compact notation in Eq.~(\ref{eq:cMPO}) denotes that $\id$ (identity operator) and $Q$ are scalars, $\bm{L}$ ($\bm{R}$) is a $(D-1)$-dimensional row (column) vector, and $\bm{P}$ is a $(D-1)$-dimensional square matrix. $Q$ is related to the local terms in the Hamiltonian, $\bm{L}, \bm{R}$ contain interaction to neighboring sites, and $\bm{P}$ is responsible for long-range interactions. The concrete form of $Q, \bm{L}, \bm{R}, \bm{P}$ can be read out from the MPO representation of the Hamiltonian, examples of which are given in Ref.~\cite{supple}. 

\begin{figure}[t]
  \resizebox{\columnwidth}{!}{\includegraphics{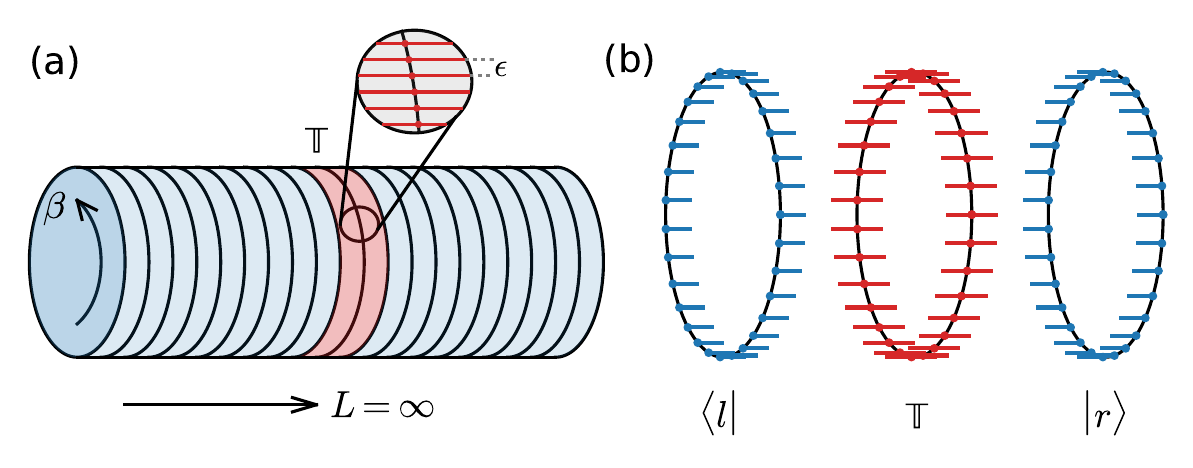}}
  \caption{(a) The partition function as a space-time tensor network living on an infinite cylinder. (b) The transfer matrix $\mathbbm{T}$ as a cMPO and its left and right dominant eigenvectors as two cMPSs.
  } \label{fig:concept}
\end{figure}

In this framework, the partition function at finite temperature is written as
\begin{equation}
Z =\tr(e^{-\beta H}) = \tr [(\cdots \manyplus \cdots )^{\beta/\dtau} ], \label{eq:PartitionFunc}
\end{equation}
where the $\beta/\dtau$-th power of the MPO (representing the infinitesimal evolution operator $e^{-\dtau H}$) indeed recovers $e^{-\beta H}$ and the trace connects the remaining physical indices pointing upwards and downwards. The partition function Eq.~(\ref{eq:PartitionFunc}) has an explicit tensor network representation formed by stacking the local tensors defined in \Eq{eq:cMPO}. In the thermodynamic limit $L \rightarrow \infty$, the tensor network has an infinite cylinder geometry as shown in Fig.~\ref{fig:concept}(a).

We highlight the central object of the present approach in Fig.~\ref{fig:concept}(a), the transfer matrix $\mathbbm{T}$, which is an MPO with horizontal open legs along the spatial direction. In the limit $\dtau \rightarrow 0$, $\mathbbm{T}$ becomes a cMPO with ``length'' $\beta$ in the periodic imaginary-time direction. 
In the thermodynamic limit, the dominant eigenvalue of $\mathbbm{T}$, which we denote by $\lambda_{\max}$ and assume to be unique~\footnote{The degeneracy in the dominant eigenvalue of $\mathbbm{T}$ indicates long-range correlations between local operators. This cannot occur at finite temperature for generic one-dimensional quantum systems with local interactions, but may happen in long-range interacting systems.}, completely determines the partition function, $Z = \lim_{L\rightarrow \infty}\lambda_{\max}^L$~\cite{Bursill1996, Wang1997}. The left and right eigenvectors associated with $\lambda_{\max}$ are respectively denoted by $\langle l |$ and $|r \rangle$ and satisfy
\begin{equation}
\mathbbm{T} |r \rangle = \lambda_{\max} |r \rangle, \quad  \langle l| \mathbbm{T} =  \langle l| \lambda_{\max}.  \label{eq:eigenvec}
\end{equation}

Under the tensor network framework, we approximate $\langle l |$ and $|r \rangle$ by using two MPSs with finite bond dimensions [see Fig.~\ref{fig:concept}(b)]. The MPSs are taken to be uniform, i.e., translationally invariant along the imaginary-time direction. For instance, the MPS for $|r\rangle$ is defined through a local tensor
\begin{equation}
\r_i = \left(\begin{array}{c}
  \id_{\r} + \dtau  Q_{\r} \\
    \sqrt{ \dtau} \bm{R}_{\r}
  \end{array}\right)_{i},  \label{eq:cmps}
\end{equation}
where $i\in[1,D]$ is the index of the horizontal tensor legs in the spatial direction.
The MPS has bond dimension $\chi$ along the imaginary-time direction. 
Hence, $\id_{\r}$ is the $\chi$-dimensional identity matrix, $Q_{\r}$ is a $\chi \times \chi$ matrix, and $\bm{R_{\r}}$ is a $(D-1)$-dimensional column vector, each entry of which is a $\chi \times \chi$ matrix. The local tensor $\l_i$ for defining $\<l|$ has the same structure as Eq.~(\ref{eq:cmps}), which contains $\id_{\l}, Q_{\l}$, and $\bm{R_{\l}}$. In the limit $\dtau \rightarrow 0 $, both the left and right MPSs reduce to cMPS. Compared to the original formulation of cMPSs in the continuous space~\cite{Verstraete2010a}, the imaginary-time direction is continuous and periodic in the present setting.

Given the left and right cMPSs, the dominant eigenvalue of $\mathbbm{T}$ can be estimated from the quotient $
\lambda_{\max} = {\langle l| \mathbbm{T} |r \rangle}/{\langle l |r\rangle}$. 
First, to compute the overlap $\langle l|r\rangle$ in the denominator, one can form a (temporal) transfer matrix $\gong$ along the imaginary-time direction, where the connected leg sums over the horizontal indices of two cMPS local tensors. To the leading order of $\dtau$, one has~\cite{Verstraete2010a}
\begin{equation}
  \gong =\id_{\l} \otimes \id_{\r}  +  \dtau \underbrace{  (   {Q}_{\l} \otimes \id_{\r} + \id_{\l} \otimes {Q}_{\r} + \bm{R}_{\l}\otimes \bm{R}_{\r} )}_{-K_{\gong}}, \label{eq:gong-mat}
\end{equation}
where $\bm{R}_{\l}\otimes \bm{R}_{\r}$ already assumed summation over the horizontal index, and $K_{\gong}$ is a $\chi^2\times \chi^2$ matrix by combining two legs upward and downward. Given \Eq{eq:gong-mat}, the overlap can be evaluated in the continuous-time limit as~\cite{Verstraete2010a}
\begin{equation}
  \langle l | r \rangle = \lim_{\dtau
  \rightarrow 0} \ensuremath{\operatorname{Tr}} \left( \gong ^{\beta /
  \dtau} \right) = \tr e^{-\beta K_{\gong}}. \label{eq:up}
\end{equation}

Moreover, applying a cMPO to a cMPS yields another cMPS, as long as the first-order terms in $\dtau$ are retained. Thus, $\mathbbm{T}|r\rangle$ can be viewed as a cMPS defined via a local tensor
\begin{equation}
\tu_i = \left(\begin{array}{c}
    \id_{\+} \otimes \id_{\r} + \dtau (\id_{\+} \otimes Q_{\r} + Q_{\+} \otimes \id_{\r} +  \bm{L}_{\+} \otimes \bm{R}_{\r})   \\
    \sqrt{ \dtau } (\bm{R}_{\+} \otimes \id_{\r} +  \bm{P}_{\+} \otimes \bm{R}_{\r})
  \end{array}\right)_{i}, \label{eq:tu}
\end{equation}
where the subscript $\+$ refers to tensors from $\mathbbm{T}$ [see Eq.~(\ref{eq:cMPO})]. Then, the overlap $\langle l|$ and $\mathbbm{T}|r\rangle$ involves another (temporal) transfer matrix $\wang =\id_{\l}\otimes\id_{\+} \otimes \id_{\r} - \dtau K_{\wang}$ defined in a similar way as \Eq{eq:gong-mat}, with $K_{\wang}$ being a $\chi^2d\times \chi^2d$ matrix. Given this, the expectation reads
\begin{equation}
  \langle l | \mathbbm{T} | r \rangle  =
  \lim_{ \dtau \rightarrow 0} \tr \left(\wang^{\beta /
 \dtau} \right) =\tr e^{-\beta K_{\wang} }.  \label{eq:dn}
\end{equation}
It is worth emphasizing that taking the $ \dtau \rightarrow 0$ limit in Eqs.~(\ref{eq:up}) and (\ref{eq:dn}) has eliminated the time-discretization error.

By combining Eqs.~(\ref{eq:up}) and (\ref{eq:dn}), the free-energy density $f = -\ln Z /(\beta L) $ takes a simple form
\begin{equation}
f = -\frac{1}{\beta} \left(\ln \tr e^{-\beta K_{\wang}} - \ln \tr e^{-\beta K_{\gong}} \right).  \label{eq:free-energy}
\end{equation}
In the cases where the spatial transfer matrix $\mathbbm{T}$ is Hermitian, one has $|l\> = |r\> $. One can directly minimize \Eq{eq:free-energy} with respect to the cMPS tensors according to the variational free-energy principle. This calculation is equivalent to optimizing a periodic uniform cMPS as the dominant eigenvector of a Hermitian cMPO. 
Without loss of generality, notice that the temporal transfer matrices $\gong$ and $\wang$ can always be gauged to be Hermitian, so are $K_{\gong}$ and $K_{\wang}$~\footnote{For example, one way to keep the Hermicity of $K_{\gong}$ and $K_{\wang}$ is to choose the gauge where all matrices in the cMPO are Hermitian, and, meanwhile, fix the matrices in the cMPS to be Hermitian during variational optimizations. In practice, we can avoid the difficulty in choosing the gauge by always manually symmetrizing $K_{\gong}$ and $K_{\wang}$ before diagonalizing them.}.
The matrix exponential can be evaluated with eigendecomposition with a computational cost of $\mathcal{O}(\chi^6)$.
It is nevertheless possible to further reduce the complexity by using an approximation scheme to compute the trace exponentials. In particular, at the zero-temperature limit $\beta \rightarrow \infty$, the trace exponentials are dominated by the smallest eigenvalues of $K_{\wang}$ and $K_{\gong}$. In this case, the scaling reduces to $\mathcal{O}(\chi^3)$ if one employs a dominant eigensolver for $K_{\wang}$ and $K_{\gong}$.

In more general cases, the spatial transfer matrix $\mathbbm{T}$ is non-Hermitian. Under such a situation, we employ the power method to find the cMPS approximations of the dominant eigenvectors $\langle l |$ and $|r \rangle$. In each step of the power projection, we apply the transfer matrix $\mathbbm{T}$ to an initial cMPS with bond dimension $\chi$ and obtain a cMPS built by $\tu$ with an enlarged bond dimension $\chi d$. We then compress it back to bond dimension $\chi$ by variationally maximizing the fidelity between the target cMPS $|\psi\>$ and $\mathbbm{T}|r\>$
\begin{equation}
\mathcal{F} = {\langle \psi | \mathbbm{T} | r \rangle}/ {\sqrt{\langle \psi | \psi\rangle}  }.
\label{eq:fidelity}
\end{equation}
Calculations involved in this objective function are similar to those of the free-energy density in \Eq{eq:free-energy}. To optimize these quantities, we perform gradient-based variational optimizations using the limited-memory Broyden-Fletcher-Goldfarb-Shanno (L-BFGS) quasi-Newton algorithm, where we employ the differentiable programming approach~\cite{Liao2019a} to compute the gradient with respect to the cMPS tensors conveniently. The explicit derivation of the gradient and the initialization strategies of the variational optimization are given in Ref.~\cite{supple}.

After obtaining the boundary cMPS approximations for the dominant eigenvectors of the spatial transfer matrix $\mathbb{T}$, one can compute a number of physical observables. First, defining $ O = \id_{\l} \otimes O_i \otimes \id_{\r} $ allows one to calculate the thermal average of local operators as
\begin{equation}
\langle O_i \rangle =  \tr \left(\mathrm{e}^{- \beta K_{\wang} } O\right)
/\tr \left(\mathrm{e}^{ -\beta K_{\wang}  }\right), \label{eq:localO}
\end{equation}
where, curiously, $K_{\wang}$ acts as an ``effective Hamiltonian'' and $e^{-\beta K_{\wang}}$ plays the role of the single-site reduced density matrix after leaving the physical indices untraced~\footnote{Computing spatially nonlocal correlation function will involve constructing the transfer matrix $\l\cdots\manyplus\cdots \r$ and contracting the resulting tensor network approximately.}. Having access to the temporal transfer matrix $\wang$ also allows computing the local two-time correlation function conveniently~\cite{Banuls2009}
\begin{equation}
\< A_i(\tau) B_i \>  = \tr\left( e^{- (\beta - \tau)K_{\wang}} A e^{ - \tau K_{\wang}}  B \right )/  \tr\left( e^{-\beta K_{\wang}} \right).
\end{equation}
An example of this is the spin-spin correlation function $\chi(\tau) \equiv \< S_i^z(\tau) S_i^z \> $. The corresponding Matsubara frequency susceptibility $\chi(i\omega)$ is computed using the eigenvalue decomposition of the effective Hamiltonian $K_{\wang}$. Crucially, one can directly perform analytic continuation to real frequencies to obtain the dynamical susceptibility $\chi''(\omega) \equiv \mathrm{Im}\, \chi(i\omega \rightarrow \omega +i0^{+})$ given the spectral representation. Moreover, the local spectral function $ S(\omega) = {2 \chi{''}(\omega)}/({1-e^{-\beta \omega}})$ follows according to the fluctuation-dissipation theorem~\cite{SachdevQPT}. The details on the computation of these dynamical quantities in the cMPO framework are given in Ref.~\cite{supple}.

Moreover, the energy density and the specific heat can be directly computed by taking the explicit derivative of the free-energy density in \Eq{eq:free-energy} with respect to the inverse temperature~\footnote{The implicit derivative through the boundary cMPS vanishes at the variational extreme, as a consequence of the Feynman-Hellmann theorem.},
\begin{align}
e &= \< K_{\wang} \>_{K_{\wang}}- \langle K_{\gong} \rangle_{K_{\gong}}, \label{eq:en} \\
c &=  \beta^2 \left[ \left(\< K_{\wang}^2 \>_{K_{\wang}} - \< K_{\wang} \>^2_{K_{\wang}} \right)  -  \left(\< K_{\gong}^2 \>_{K_{\gong}} - \< K_{\gong} \>^2_{K_{\gong}} \right) \right], \label{eq:cv}
\end{align}
where the notations $\< \,\cdot \, \>_{K_{\gong}}$ and $\<\, \cdot \,\>_{K_{\wang}}$ stand for the thermal average over the effective Hamiltonians $K_{\gong}$ and $K_{\wang}$, respectively. Having a direct estimator for the specific heat is more convenient than computing numerical differentiation on a fine scan of temperature. However, one should also be cautioned that the estimators in Eqs.~(\ref{eq:en}) and (\ref{eq:cv}) hold only upon the convergence of the power projection. The specific heat estimator can be less reliable than numerical differentiation at low temperature since it involves subtraction of large numbers, similar to the case of stochastic series expansion QMC method~\cite{Sandvik2010a}.
Our code implementation is publicly available at Ref.~\footnote{See \url{https://github.com/TensorBFS/cMPO} for code implementation in PyTorch}.

{\em Results} --- As the first application, we consider the quantum spin-1/2 $XXZ$ chain
\begin{equation}
 H = \sum_{ \<i, j\>} (S_i^x S_j^x + S_i^y S_j^y + \Delta  S_i^z
   S_j^z ), \label{eq:XXZ}
\end{equation}
where $\Delta$ is the anisotropy parameter. The model reduces to the quantum $XY$ model at $\Delta=0$ and the Heisenberg model at $\Delta=1$. In general, the cMPO representation for the quantum $XXZ$ chain has $D=4$~\cite{supple}. We note that, when $\Delta\geq0$, it is possible to perform a basis rotation to bring the cMPO into a Hermitian form which allows a direct variational optimization of \Eq{eq:free-energy}. We nevertheless employ the power method for its generality. The bond dimension of the boundary cMPS is fixed to be $\chi=20$ in this study.

\begin{figure}[t]
   \includegraphics[width=\columnwidth, trim={1cm 1cm 2cm 2cm}, clip]{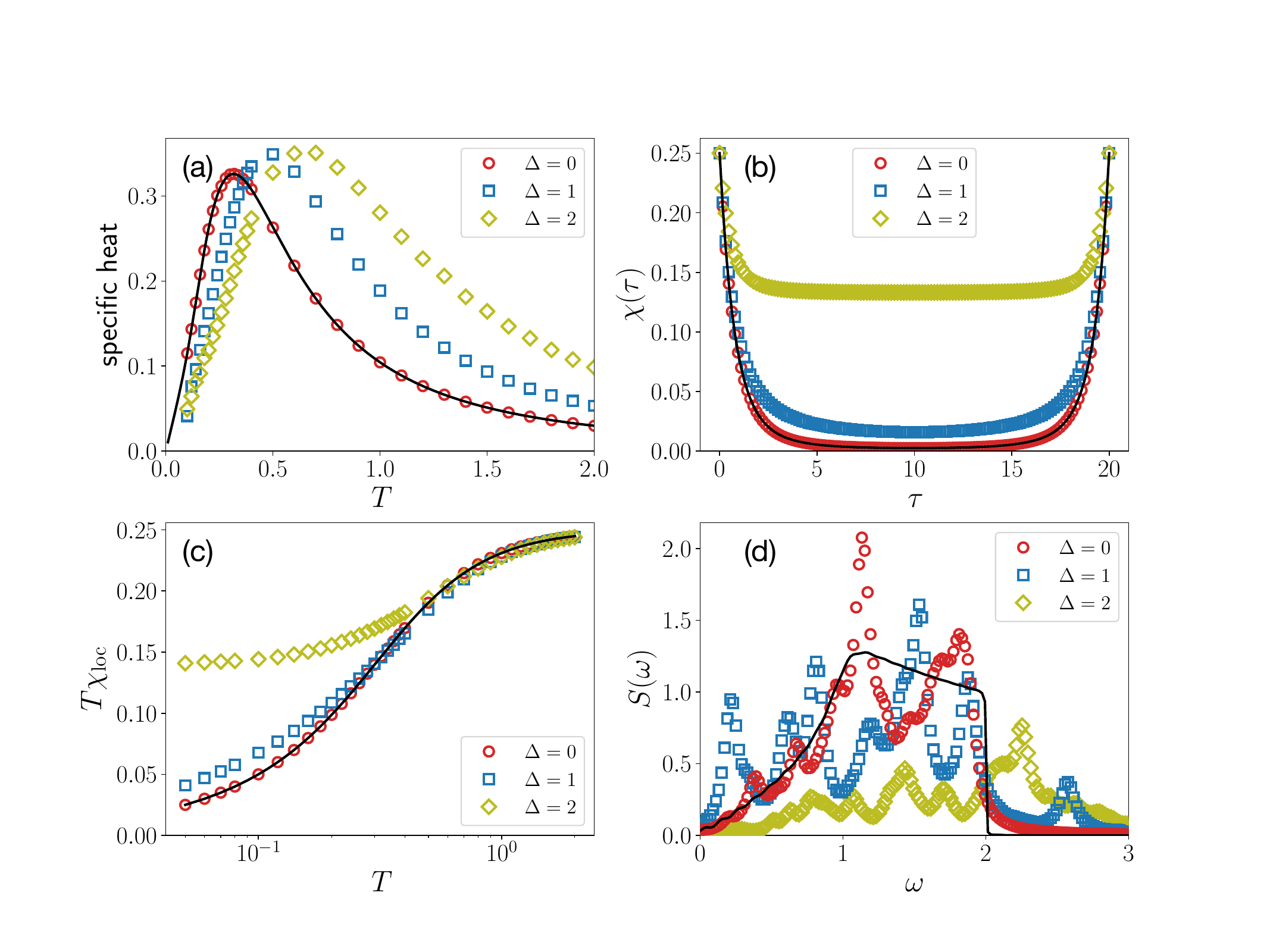}
  \caption{(a) The specific heat computed with \Eq{eq:cv}; (b) local unequal time spin correlation function at $\beta=20$; (c) the local susceptibility; (d) the local spectral function of the $XXZ$ chain with various anisotropies. The solid black lines are the exact results in the $XY$ limit.
  \label{fig:xxz}
  }
\end{figure}

Figure~\ref{fig:xxz}(a) shows the specific heat of the quantum $XXZ$ chain at various $\Delta$. For both $XY$ and Heisenberg cases, the specific heat vanishes linearly with respect to $T$, which is in agreement with conformal field theory predictions~\cite{affleck_universal_1986,blote_conformal_1986}.
Figure~\ref{fig:xxz}(b) shows that the imaginary-time correlator reaches higher values in the large time limit with increasing of $\Delta$, indicating the development of long-range correlations along the imaginary-time direction. As a consequence, the local susceptibility $T \chi_\mathrm{loc} = \frac{1}{\beta}\int_0^\beta d\tau \chi(\tau)$ indicates the local moment in the low-temperature limit as shown in Fig.~\ref{fig:xxz}(c).
In the $XY$ case, all of the physical quantities computed with the cMPO method are in excellent agreement with the exact results~\cite{supple}.

Furthermore, Fig.~\ref{fig:xxz}(d) shows the local spectral function $S(\omega)$ calculated directly in the real frequency using the spectral decomposition of $K_{\wang}$. Although dynamical properties are more sensitive to the convergence of the boundary cMPS than thermodynamic quantities, the comparison to the exact results in the $XY$ limit~\cite{Cruz1981} shows encouraging agreement, especially in the low-frequency region. 
In particular, the zero frequency value of the local spectral function is related to the nuclear magnetic resonance (NMR) spin-lattice relaxation rate measured in experiments~\cite{Moriya1962} which indicates the strength of low-energy fluctuations. There have been extensive efforts in studying this quantity both analytically and numerically~\cite{Sachdev1994, Steinberg2019, Sandvik1995, Starykh1997, Dupont2016a, Coira2016, Capponi2019}. 
At high frequencies, the multipeak structure and the inconsistency with the exact solution are attributed to the finiteness of the ``effective Hamiltonian.'' The high-frequency results can be improved by increasing the bond dimension.
By far, two predominant numerical approaches are based on analytic continuation of imaginary time correlation functions~\cite{Jarrell1996, Sandvik1998, wang1999calculation} and Fourier transform of the extrapolated real-time data~\cite{sirker2005real, Sirker2006, Barthel2009, Karrasch2012}. The cMPO approach offers a way to compute finite temperature spectral functions without getting into the tricky business of analytic continuation of imaginary-time data or prediction of real-time series. Moreover, the cMPO approach applies more broadly to frustrated systems with long-range interactions, which means it is applicable to
quasi-one-dimensional systems with a cylindrical geometry, in a similar spirit as the applications of the DMRG in this geometry~\cite{stoudenmire2012studying}. Further investigations are needed to fully explore this direction and make an extensive and quantitative comparison between various approaches. We remark that, as a bottom line, one can always only deal with the imaginary-time data~\cite{Mutou1998,Naef1999} and employ recent advances in analytic continuation to impose prior knowledge in the spectrum~\cite{Sandvik2016, Shao2017a}.

\begin{figure}[!t]
  \includegraphics[width=0.45\columnwidth, trim={0cm 0cm 0cm 0cm}, clip]{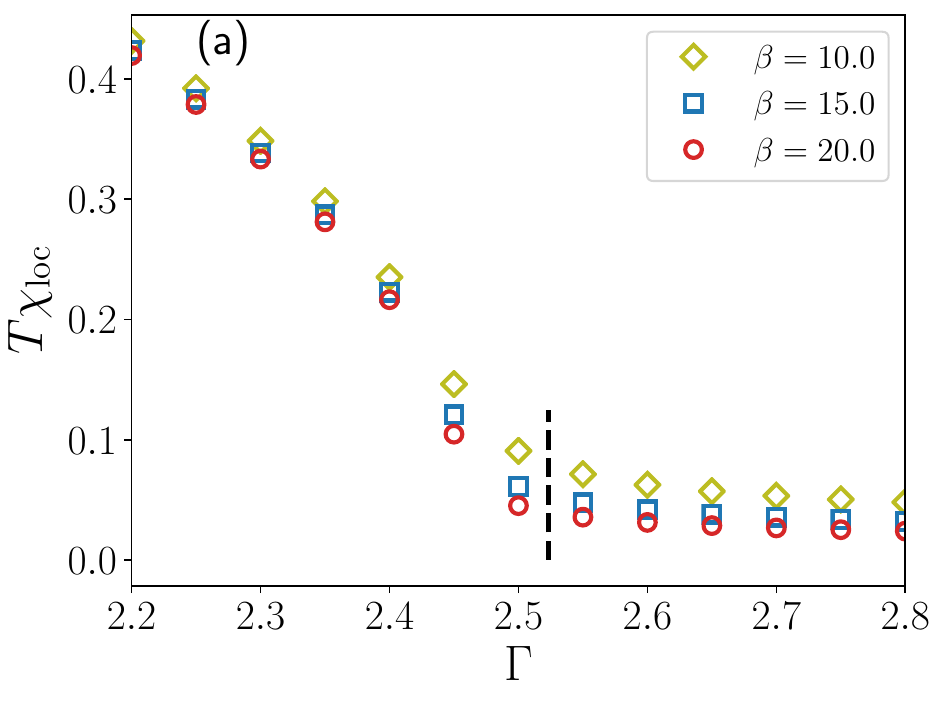}
  \includegraphics[width=0.5\columnwidth, trim={0cm 0cm 0cm 1cm}, clip]{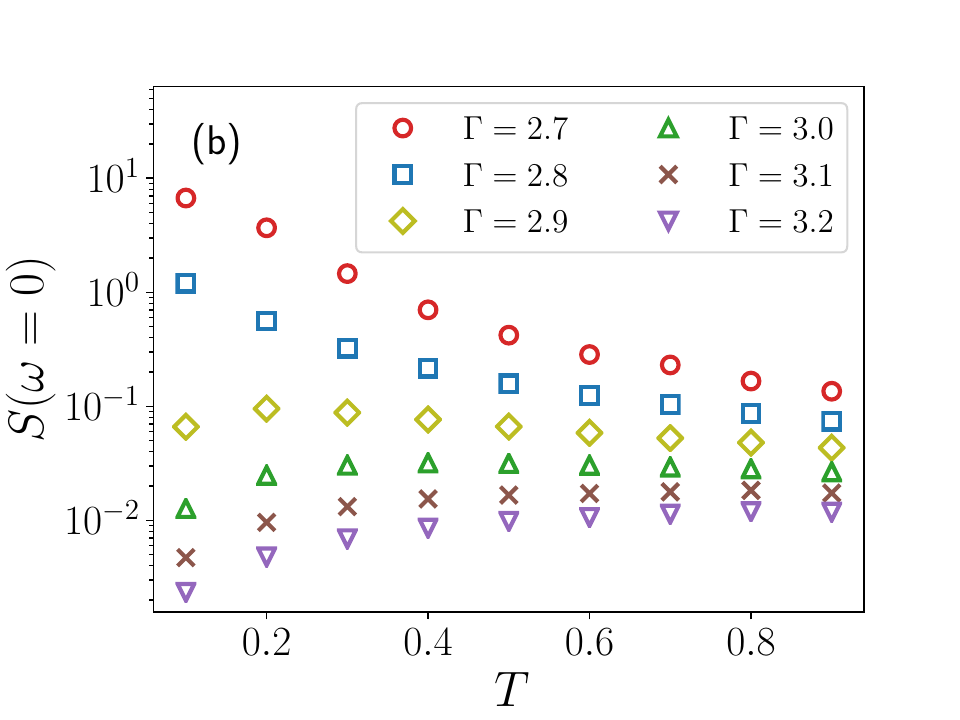}
  \caption{(a) The local susceptibility of the inverse-square Ising chain with a transverse field. The bond dimension of the cMPS is $\chi=20$. The vertical dashed line marks the quantum multicritical point at $\Gamma_c\approx2.5236$~\cite{Fukui2009, syngemonte}.
   (b) The zero frequency local spectral function of the transverse field Ising model on a cylinder with width $W=4$.
  \label{fig:longrangeising}
  }
\end{figure}

Finally, we consider the transverse field Ising model with long-range interactions defined by the following Hamiltonian:
\begin{equation}
  H = - \sum_{i < j} J_{i,j} Z_i Z_j - \Gamma \sum_i X_i,  \label{eq:long-range-ising}
\end{equation}
where $X$ and $Z$ are Pauli matrices. First, assuming the spins are arranged in a one-dimensional chain and the coupling follows a power-law decaying interaction $J_{i,j} ={J}/{| j - i |^{\alpha}}$, the Hamiltonian is relevant to trapped ions and Rydberg atom quantum simulators realized experimentally~\cite{Britton2012, Richerme2014, Bohnet2016, Zeiher2017}. Although the power $\alpha$ is tunable in a range, we focus here on the case of inverse-square interaction, i.e., $\alpha =2$, where the model exhibits a Kosterlitz-Thouless transition as temperature changes~\cite{Anderson1971,Kosterlitz1973}. When $J=1$, in the ground state the model exhibit a quantum multicritical point at $\Gamma_c\approx2.5236$~\cite{Fukui2009, syngemonte}. To handle the long-range interaction, we follow Ref.~\cite{Crosswhite2008} to represent the power-law decaying interaction as a sum of exponentials~\cite{supple}.
Figure~\ref{fig:longrangeising}(a) shows the local susceptibility calculated in the inverse-square Ising chain.
In the ordered phase, the local susceptibility remains a finite value at zero temperature, while it vanishes in the disordered phase.
By calculating $T\chi_{\mathrm{loc}}$ for different transverse fields and temperatures, we find our results consistent with the QMC prediction on the location of the quantum multicritical point~\cite{Fukui2009, syngemonte}. 

Next, we turn to the model in \Eq{eq:long-range-ising} on a quasi-one-dimensional lattice with cylindrical geometry. Consider an infinitely long cylinder with circumference $W$. If we impose the helical boundary condition, the Hamiltonian can be regarded as a one-dimensional Ising chain with $J_{i,i+j}=J$ if $j=1 \text{ or } W$, and zero otherwise. The bond dimension of the cMPO is $D=W+1$~\cite{supple}. In the two-dimensional limit $W\rightarrow \infty$, the transverse field Ising model shows a quantum critical point at $\Gamma_c=3.04438(2)$~\cite{Rieger1999, PhysRevE.66.066110}. Here we focus on a finite width cylinder $W=4$ and study the temperature dependence of zero-frequency local spectral function across the quantum critical point. Figure \ref{fig:longrangeising}(b) shows that in the ordered phase the spectral function increases when lowering the temperature due to the presence of elastic modes. Meanwhile, the spectral function is suppressed in the quantum disorder phase at the low-temperature limit. Such drastically different behaviors were observed previously in the NMR relaxation rate across the quantum critical point of the quantum Ising chain~\cite{Kinross2014}. Having a numerical probe of the quantity for the quasi-one-dimensional case opens an opportunity to access the spectral information of frustrated magnets and even fermions in two dimensions~\cite{Yoshitake2016, Jansa2018}.



{\em Summary and discussion} --- To summarize, we have put forward an algorithm for studying quantum systems at finite temperature by developing cMPO techniques. This approach works directly in the thermodynamic limit and does not have any time discretization error. Moreover, this approach works well for long-range interactions and can compute local dynamical properties. The basics of the present approach is the MPO representation of the evolution operator in the continuous-time limit. For future works, the compression scheme for MPO representation of long-range interacting Hamiltonians~\cite{Chan2016, Stoudenmire2017a, Hubig2017} may be used to obtain more compact cMPOs for thermal state calculations. In the cMPO framework, the effective Hamiltonians $K_{\gong}$ and $K_{\wang}$ play a central role in the computation as well as governing the physics. It would be worth investigating their properties and universal behaviors more closely.

The cMPO representation can also be extended to higher-dimensional tensor networks~\cite{supple}. Given the contraction algorithm of the partition function in the 1+1 dimensions presented here, one envisions a contraction scheme in the 2+1 dimensions along the line of Refs.~\cite{Nishino2000, Okunishi2001, Vanderstraeten2018a}, which would directly provide thermal properties of infinite two-dimensional systems other than the cylinders considered here. However, this algorithm also features nested variational optimization and projection with higher computational costs, which deserve further investigations~\cite{Tang-unpublished}.

{\em Acknowledgment} --- We thank Tao Xiang, Hai-Jun Liao, Zhi-Yuan Xie, and Wei Li for fruitful discussions. L.~W.~is supported by the Ministry of Science and Technology of China under the Grants No.~2016YFA0300603 and 2016YFA0302400, the National Natural Science Foundation of China under Grant No.~11774398.
H.-H.~T.~is supported by the DFG through project A06 of SFB 1143 (project-id 247310070).

\bibliography{cmpo-arxiv}

\bibliographystyle{apsrev4-2}

\clearpage

\section*{Supplemental Material}

\setcounter{table}{0}
\renewcommand{\thetable}{S\arabic{table}}
\setcounter{figure}{0}
\renewcommand{\thefigure}{S\arabic{figure}}
\setcounter{equation}{0}
\renewcommand{\theequation}{S\arabic{equation}}

\appendix 
\section{A cMPO cookbook} \label{appendix:cmpo-zoo}
We compile a catalog of cMPOs for frequently encountered models. One can verify these construction by multiplying local tensors in the form of $\cdots \manyplus \cdots$ and restores $e^{-\dtau H} $ to the leading order of $\dtau$.

\subsection{Transverse field Ising chain}

The Hamiltonian of transverse field Ising chain reads
\begin{equation}
H = - \Gamma \sum_i X_i - J \sum_{\langle i, j \rangle} Z_i Z_j.
\end{equation}
The corresponding cMPO tensor reads
\begin{equation}
\+ = \left(\begin{array}{cc}
    \id +  \dtau   \Gamma X & \sqrt{\dtau J} Z\\
    \sqrt{\dtau J}  Z & 
  \end{array}\right) \label{eq:tfim-cmpo}
\end{equation}
with physical dimension $d=2$ and virtual bond dimension $D=2$. 
One can read out the sub tensors $Q, \bm{R}, \bm{L}$ and $\bm{P}$.
Note that the spatial transfer matrix is Hermitian for ferromagnetic couplings $J>0$. 
In that case $|r\> = |l\>$ and one can variationally optimize the free energy in Eq.~(\ref{eq:free-energy}) directly.

\subsection{XXZ chain}
The Hamiltonian is given by 
 \begin{equation}
 H = \sum_{ \<i, j\>} (S_i^x S_j^x + S_i^y S_j^y + \Delta  S_i^z
   S_j^z ), \label{eq:xxz-chain}
 \end{equation}
where $S_i^{\alpha}$ ($\alpha = x, y, z$) are spin-$1 / 2$ operators. 
Introducing $S^\pm \equiv S^x \pm i S^y$, the cMPO tensor is given by
\begin{equation}
 \+ = \left(\begin{array}{c|ccc}
     \id & \sqrt{ \dtau} S^+ & \sqrt{\dtau}  S ^- &  \sqrt{\dtau} S^z\\ \hline
    - \sqrt{\dtau} S^-/2 &  &  & \\  
    - \sqrt{\dtau} S^+/2 &  &  & \\
    - \Delta \sqrt{\dtau} S^z &  &  & 
     \end{array}\right) . \label{eq:xxz-cmpo} 
\end{equation}   
The cMPO has physical bond dimension $d=2$ and virtual bond dimension $D=4$.
   
\subsection{3-state Potts chain}
The Hamiltonian is given by
\begin{equation}
 H = - J \sum_{\langle i, j \rangle} (\sigma_i \sigma_j^2 + \sigma_i^2
   \sigma_j) - \Gamma \sum_i (\tau_i + \tau_i^2) .
\end{equation}
where
\begin{equation}
 \sigma = \left( \begin{array}{ccc}
     1 &  & \\
     & \omega & \\
     &  & \omega^2
   \end{array} \right), \tau = \left( \begin{array}{ccc}
     & 1 & \\
     &  & 1\\
     1 &  & 
   \end{array} \right) .
\end{equation}
Here $\omega = \exp (2 \pi \mathrm{i} / 3)$. The cMPO tensor is given by 
\begin{equation}
 \+ = \left( \begin{array}{c|cc}
     \id +  \dtau \Gamma (\tau + \tau^2) & \sqrt{\dtau} \sigma &
     \sqrt{\dtau} \mathrm{} \sigma^2\\
     \hline
     J \sqrt{\dtau} \sigma^2 &  & \\
     J \sqrt{\dtau} \sigma &  & 
   \end{array} \right) . 
\end{equation}
The cMPO has physical bond dimension $d=3$ and virtual bond dimension $D=3$.
   
\subsection{Bose-Hubbard Model}
The Hamiltonian is given by
\begin{equation} 
 H = - J \sum_{\langle i, j \rangle} (b_i^{\dagger} b_j + b_i b_j^{\dagger})
   - \mu \sum_i n_i + \frac{1}{2} U \sum_i n_i  (n_i - 1),   
\end{equation} 
where $n = b^{\dagger} b$. The matrix forms of $b$ and $b^{\dagger}$ are given
by
\begin{equation}
 b = \left( \begin{array}{ccccc}
     0 & 1 &  &  & \\
     & 0 & \sqrt{2} &  & \\
     &  & 0 & \sqrt{3} & \\
     &  &  & 0 & \ddots\\
     &  &  &  & \ddots
   \end{array} \right), b^{\dagger} = \left( \begin{array}{ccccc}
     0 &  &  &  & \\
     1 & 0 &  &  & \\
     & \sqrt{2} & 0 &  & \\
     &  & \sqrt{3} & 0 & \\
     &  &  & \ddots & \ddots
   \end{array} \right) .
\end{equation}
The cMPO tensor reads
\begin{equation}
 \+ = \left( \begin{array}{c|cc}
     \id +  \dtau (\mu n - Un (n - 1) / 2) & \sqrt{ \dtau} b^\dagger & \sqrt{  \dtau} b \\ \hline
   J  \sqrt{ \dtau} b &  & \\
   J  \sqrt{  \dtau} b^\dagger &  & 
   \end{array} \right) . 
\end{equation}
The physical bond dimension is given by the truncation of the boson occupation number. And the
virtual bond dimension is $D = 3$.

\subsection{Quantum Ising chain with long-range interactions}

The Hamiltonian reads 
\begin{equation}
  H = - \sum_{i < j} \frac{J}{| j - i |^{\alpha}} Z_i Z_j - \Gamma \sum_i
  X_i. 
\end{equation}

After decomposing the power-law decaying interaction into a sum of exponentials~\cite{Crosswhite2008},
$
  \frac{J}{| j-i |^{\alpha}} = \sum_{k = 1}^K \mu_k e^{- \lambda_k (j -
  i)}
$, the cMPO tensor reads
  
\begin{equation}
 \+ = \left(\begin{array}{c|cccc}
    \id + \dtau \Gamma X & \sqrt{\dtau}  Z & \sqrt{\dtau} Z & \ldots & \sqrt{\dtau}  Z\\ \hline
  \sqrt{\dtau}  \mu_1 e^{- \lambda_1} Z & e^{- \lambda_1} \id &  &  & \\
  \sqrt{\dtau}   \mu_2  e^{- \lambda_2} Z &  & e^{- \lambda_2} \id &  & \\
    \vdots &  &  & \ddots & \\
   \sqrt{\dtau}  \mu_K  e^{- \lambda_K} Z &  &  &  & e^{- \lambda_K} \id
  \end{array}\right). 
\end{equation}
The virtual bond dimension of the cMPO representation is determined by the number of exponentials, $D=K+1$.

\subsection{$J_1$-$J_2$ Heisenberg chain}

The Hamiltonian is defined on a chain with nearest and next-nearest neighbor interactions
\begin{equation}
H = \sum_{\langle i, j \rangle}  J_1 \bm{S}_i \cdot \bm{S}_j+ \sum_{\langle\langle i, j \rangle\rangle} J_2 \bm{S}_i \cdot \bm{S}_j. 
\end{equation}

The cMPO tensor has virtual bond dimension $D=7$
\begin{equation}
\+ = \left(\begin{array}{l|cccccc}
    \id & S^+ &   & S^- &  & S^z & \\ \hline
    -\sqrt{ \dtau} {J_1} S^-/2 &  & \id &  &  &  & \\ 
    -\sqrt{ \dtau} {J_2} S^-/2 &  &  &  &  &  & \\
    -\sqrt{ \dtau} {J_1} S^+/2 &  &  &  & \id &  & \\
    -\sqrt{ \dtau} {J_2} S^+/2 &  &  &  &  &  & \\
    -\sqrt{ \dtau} J_1 S^z &  &  &  &  &  & \id\\
    -\sqrt{ \dtau} J_2 S^z &  &  &  &  &  & 
  \end{array}\right).
\end{equation}

\subsection{Two-dimensional transverse field Ising model }
 
Unwrapping the cylinder into a one-dimensional chain we will get long-range interaction up to the range of the cylinder width $W$. One can directly write down the MPO representation of its Hamiltonian representation~ \cite{Crosswhite2008a, Crosswhite2008, Frowis2010}. Then, the cMPO representation of the evolution operator automatically follows. 

For simplicity we assume the helical boundary condition around 
the cylinder, the system is equivalent to a one-dimensional system with nearest neighbor and $W$-th neighbor 
interaction. Therefore, the virtual bond dimension is $D=W+1$. For example, in the case $W=4$, the cMPO tensor reads
\begin{equation}
\+ = \left(\begin{array}{l|cccc}
    \id + \dtau \Gamma X & \sqrt{\dtau} Z &  &  & \\ \hline
    \sqrt{\dtau} J Z     & & \id &  & \\
                    &  &  & \id & \\
                    &  &  &   & \id \\
     \sqrt{\dtau} J Z &  &  &  & 
  \end{array}\right).
\end{equation}
For periodic boundary condition one will have the cMPO of the same size. Except that along the one-dimensional 
system there is an unit cell of length $W$. Every first and the $W$-th cMPO of the unit cell needs to be modified~\cite{Motruk2016}. 

\section{Trotterization and worldline picture of cMPO} \label{appendix:trotter}

To gain a more intuitive understanding of the cMPO representation, we 
derive it from the continuous-time limit of the Trotter decomposition. 
Moreover, we draw a connection of the cMPO representation to the continuous-time worldline picture commonly used in quantum Monte Carlo simulations.

\begin{figure}[!htb]
  \resizebox{\columnwidth}{!}{\includegraphics{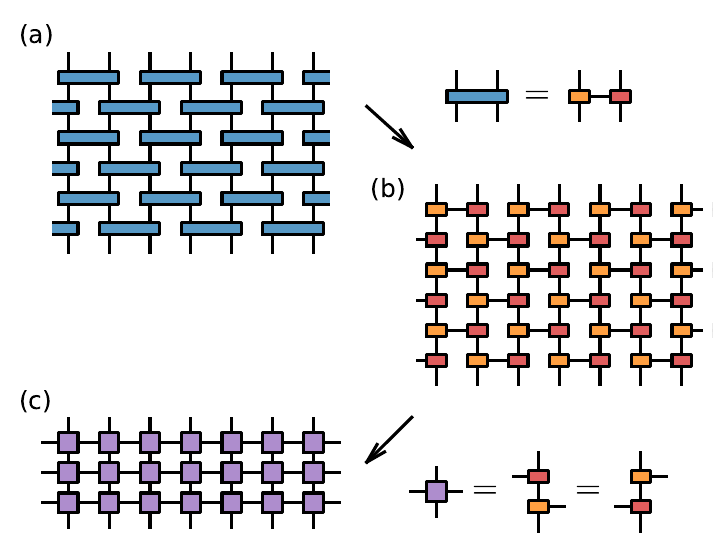}}
  \caption{Steps to obtain a cMPO representation via Trotterization. 
  (a) The tensor network constituted by Trotter gates.
  (b) The tensor network constituted by three-leg tensors $\l$ and $\r$.
  (c) The final uniform tensor network composed of cMPO.  \label{fig:trotter}}
\end{figure}

Consider a one-dimensional Hamiltonian with nearest-neighbor interactions, we perform the Trotter decomposition 
according to the Hamiltonian terms acting on even and odd bonds of the chain. 
The partition function is a tensor network constituted by the Trotter gates $\mathrm{e}^{- \dtau h_{i j}}$, see Fig.~\ref{fig:trotter}(a).
It is then conventional to perform a numerical singular-value decomposition (SVD) to the Trotter gate in order to decompose it to two three-leg tensors shown in Fig.~\ref{fig:trotter}(b). 
However, during the numerical SVD, the information of $\dtau$ is immersed into tensors and can no longer be separated. 
Alternatively, we keep track of $\dtau$ analytically and take the continuous-time limit $\dtau \rightarrow 0$ at the end of the calculation. 
This allows us to eliminate the Trotter error at the algorithmic level and gives the cMPO representation.

As a concrete example, we consider the transverse field Ising model. Recall that 
the Hamiltonian is a sum over terms on bond, i.e., $H=\sum_{\langle i, j \rangle} h_{ij}$ with
\begin{equation}
   h_{i j} = - \frac{\Gamma}{2} (X_i + X_j) - J Z_i Z_j. 
\end{equation}
Expand the evolution operator to the leading order of $\dtau$, we have
\begin{equation}
  \mathrm{e}^{- \dtau h_{i j}} 
= \id \otimes \id + \frac{\Gamma \dtau}{2} (X \otimes \id + \id
  \otimes X) + J \dtau Z \otimes Z.  
\end{equation}
Next, we write $\mathrm{e}^{- \dtau h_{i j}} = \gong$, where $\gong$ consists of two three-leg tensors
\begin{align}
  \l & = \left( \begin{array}{cc}
    \id + \frac{\Gamma}{2} \dtau X & \sqrt{\dtau J} Z
  \end{array} \right), \\
  \r & = \left( \begin{array}{c}
    \id + \frac{\Gamma}{2} \dtau X\\
     \sqrt{\dtau J} Z
  \end{array} \right) . 
\end{align}
This effectively accomplishes the same job of the original SVD process but keeps track of $\dtau$ analytically. Finally, we contract the three leg tensors vertically as shown in
Fig.~\ref{fig:trotter}(c) while keeping the leading order terms in $\dtau$. The resulting four-leg tensor
\begin{equation}
  \+ = \left( \begin{array}{cc}
    \id + \Gamma \dtau X & \sqrt{\dtau J} Z\\
    \sqrt{\dtau J} Z & 
  \end{array} \right) \label{eq:cmpo-example-ising}
\end{equation}
is the cMPO tensor for the transverse field Ising model. 
This Trotterization scheme can be generalized to higher dimensions to obtain continuous tensor network operetor (cTNO) representations.

\begin{figure}[!htb]
   \includegraphics[width=\columnwidth, trim={6cm 6cm 7cm 5cm}, clip]{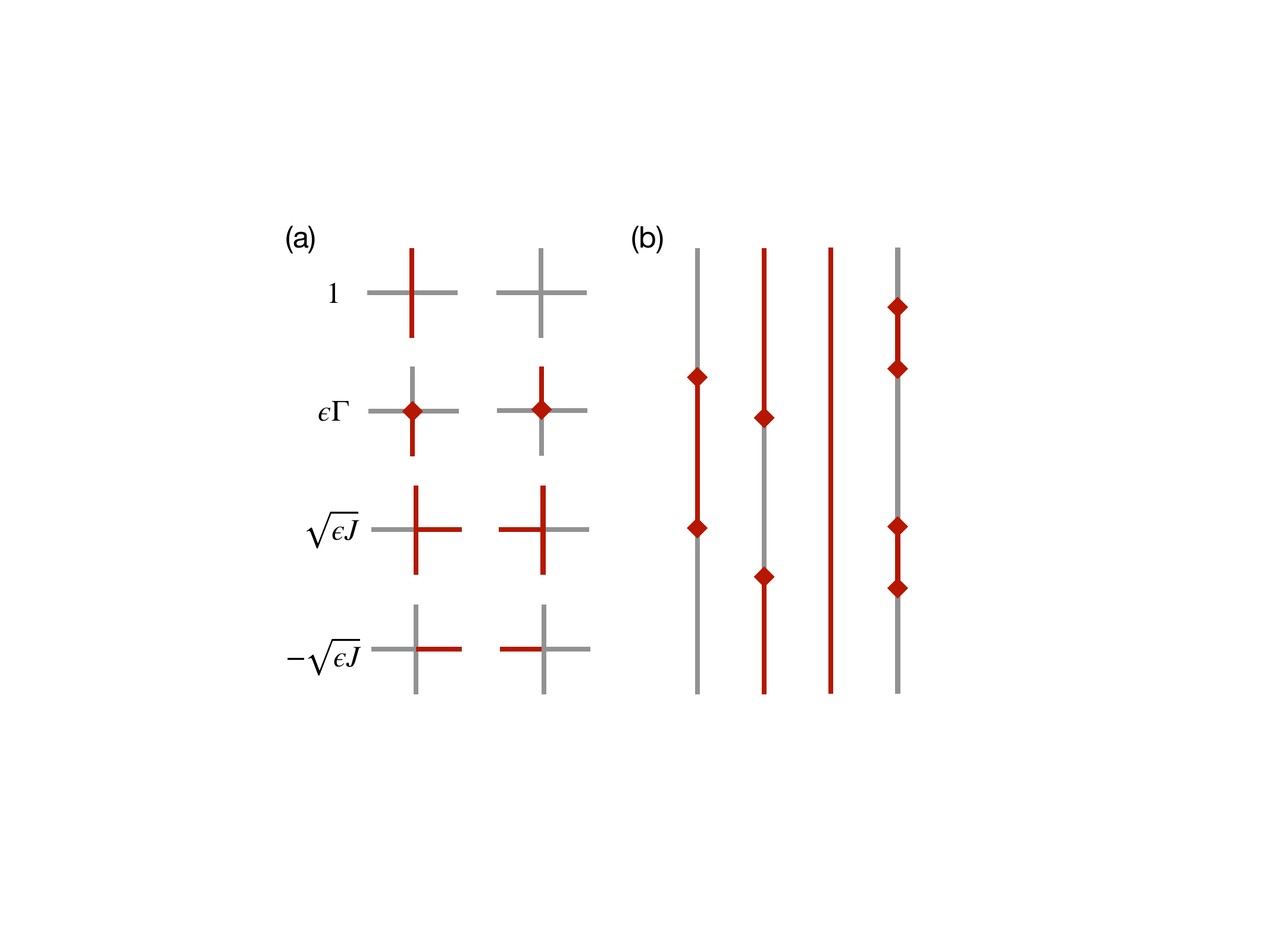}
  \caption{(a) The nonzero elements of the transverse field Ising model cMPO tensor~\Eq{eq:cmpo-example-ising}.
  (b) Continuous time worldline formed by the local cMPO tensors.  
  \label{fig:worldline}}
\end{figure}
The local tensors of the transverse field Ising model \Eq{eq:cmpo-example-ising} has the bond dimensions $d=2$, and $D=2$. 
There are only $8$ nonzeros tensor elements. Figure~\ref{fig:worldline}(a) illustrates these nonzero tensor elements using gray/red legs to denote tensor indices $0/1$.
Each tensor carries a statistical weight given by the tensor element indicated on the left. 
The first row of configurations account for the zeroth-order contribution, i.e, the identity $\id$ operator. 
The two configurations in the second row account for the on-site spin flip. We illustrated it using a diamond symbol $\vardiamondsuit$ in the center of the tensor. The last two rows represent the Ising coupling with the virtual leg index  $1$. Although the cluster expansion~\cite{Zaletel2015, Vanhecke2019} shows that these configurations only contain interactions on non-overlapping bonds, it becomes exact as $\dtau$ approaches zero. 
Amusingly, one obtains the worldline configuration shown in Fig.~\ref{fig:worldline}(b) by assembling these local tensors into a tensor network. They are precisely those configurations that the continuous-time quantum Monte Carlo (QMC) methods are designed to sample~\cite{Rieger1999}. 

This pictorial interpretation of the continuous-time tensor network holds more broadly for general models and in higher dimensions. The virtual indices indicate the presence and type of the interactions. Moreover, the lower right block $\bm{P}$ of the cMPO in Eq.~(\ref{eq:cMPO}) mediates longer-ranged interactions beyond nearest neighbors. One can directly construct the cTNO representation based on the worldline rules used in QMC calculations. For example, the partition function of two-dimensional quantum Ising model can be constructed using cTNO with physical leg $d=2$ and four virtual legs with $D=2$. Contraction of such tensor networks formed by cTNOs is an alternative way to study the thermodynamics of higher dimensional quantum systems. 

\section{Analytical gradient of the objective function} \label{appendix:gradient}
In the cMPS formalism, in order to perform the variational optimizations, we
usually need to compute the gradients of objective functions, such as the free
energy density in Eq.~(\ref{eq:free-energy}) and the fidelity in Eq.~(\ref{eq:fidelity}). Consider gradient with 
respect to the tensors in $\r$, we have, 
\begin{equation}
\partial_{\r} \ln \tr \mathrm{e}^{-\beta K_{\gong}} 
   =  -\beta \<  \partial_{\r} K_{\gong}   \>_{K_{\gong}}  . \label{eq:gradient}
\end{equation}   
   Recall that $K_{\gong} = - (Q_{\l} \otimes \id_{\r} + \id_{\l} \otimes
Q_{\r} + \ensuremath{\boldsymbol{R}}_{\l} \otimes
\ensuremath{\boldsymbol{R}}_{\r} )$. Thus, we can substitute
\begin{align*}
  \partial_{Q_{\r}} K_{\gong}& =  -\id_{\l} \otimes \Box,\\
  \partial_{\ensuremath{\boldsymbol{R}}_{\r}} K_{\gong}& = 
  -\ensuremath{\boldsymbol{R}}_{\l} \otimes \Box 
\end{align*}
into \Eq{eq:gradient} to obtain the gradient with respect to $Q$ and $\bm{R}_{\r}$ respectively.
Here, the notation $\Box$ represents the cavity in the tensor network after removing the tensor.
The gradient $\partial_{\r}
\ln \tr \mathrm{e}^{-\beta K_{\wang}}$ can be calculated in a similar way. 

\section{Initialization for the variational optimizations} \label{appendix:initialstate}

In the MPS-based algorithms such as the density-matrix renormalization group (DMRG), it is common to use open-boundary non-uniform MPS as the variational ansatz to exploit its power in optimization. 
In contrast, in our cMPO-cMPS formalism, the cMPS has the periodic boundary condition in the imaginary-time direction.
This formalism brings a concise and consistent framework and avoids the appearance of the Trotter error. 
However, it also leads to the disadvantage that, the optimization of such periodic uniform cMPS is less convenient than the conventional cases, and even the cases with periodic uniform MPS~\cite{PhysRevLett.121.230402} and infinite cMPS~\cite{PhysRevLett.118.220402}. 
Although there is no general solution to this issue, in the following, we try to reduce this disadvantage by choosing a suitable initial guess for $|\psi\>$ that is anticipated to be close enough to the global minimum.

First of all, we notice that the cMPO at the boundary site of an open-boundary system is also a cMPS since it has only one virtual leg direction. 
As a starting point for all the calculations, we project the transfer matrix $\mathbbm{T}$ onto this boundary-site cMPS for several times.
During this process, we do not truncate this state, unless the resulting state exceeds the target bond dimension $\chi$ (the protocol for this truncation is discussed below).
For the systems with very short correlation lengths (such as high-temperature systems or cMPO with very large energy gap), this ``opened'' state already gives a good estimation to the eigenstate and leads to very satisfactory results.
For more complicated systems, we need to further optimize this state. 
For systems with Hermitian transfer matrices, this ``opened'' state then serves as a good initial guess for $|\psi \>$ to variationally minimize the free-energy density.
If the transfer matrix $\mathbbm{T}$ is not Hermitian, we employ the power method starting from this ``opened'' state, and variationally truncate the cMPS to the target bond dimension $\chi$ after each power step.
For simulations with large bond dimensions, we can also start the power-method process from a optimized boundary cMPS with a smaller bond dimension.

For each variational truncation problem during the power process, a good way to initialize $|\psi \>$ is to first find an isometry $U$ that optimally projects $\tu$ back to $\r$ at a given bond dimension. 
This could be done by optimizing the overlap with the iterative SVD update~\cite{PhysRevB.79.144108}. For which we also employ the recently developed differentiable programming approach~\cite{Chen2019m}. 
Note that for infinite systems this way of truncation would correspond to the TEBD method~\cite{Orus2008}. 

However, this iterative SVD update approach also has the issue of choosing the initial isometry.
To determine the initial isometry, we construct the reduced density matrix along the time direction $\rho = \tr_{\l} \mathrm{e}^{-\beta K_{\wang}}$, where the partial trace $\tr_{\l}$ represents the trace over indices corresponding to the cMPS $\l$ (Note that here $\l$ is a ``transpose'' of $\r$, and not to confuse it as the local tensor of the left eigenvector). 
By diagonalizing $\rho$ and keep the $\chi$ largest weights, one can obtain a good initial isometry.

Overall, we found that the variational update can always improve upon the fidelity of the density matrix projection and the iterative SVD updates~\cite{Vanhecke2020}.

\section{Unequal-time correlator and dynamical susceptibility} \label{appendix:susceptibility}

The following  are standard textbook derivations except that the matrix $K_{\wang}$ plays the role of a Hamiltonian. 

First, recall that the local spin correlator $\chi(\tau) \equiv  \<S^z_i (\tau) S^z_i \> $ and $\chi(i\omega) = \int_0^\beta d\tau \chi(\tau) e^{i\omega \tau}$. Given the eigendecomposition of effective Hamiltonian $K_{\wang} = U \Lambda U^\dagger$, the Matsubara frequency susceptibility is calculated as 
\begin{equation}
\chi(i\omega) = \frac{1}{Z}\sum_{m,n}  {\tilde{S}^z}_{nm}   {\tilde{S}^z}_{mn}   \frac{e^ {-\beta \Lambda_m} - e^ {-\beta \Lambda_n} }{ i\omega - \Lambda_m + \Lambda_n  }, \label{eq:chi_matsubara}
\end{equation}
where $Z = \sum_n e^{-\beta \Lambda_n}$ and $\tilde{S}^z = U^\dagger (\id_{\l} \otimes S^z  \otimes \id_{\r} )U$. 

\begin{figure}[!htb]
  \resizebox{\columnwidth}{!}{\includegraphics{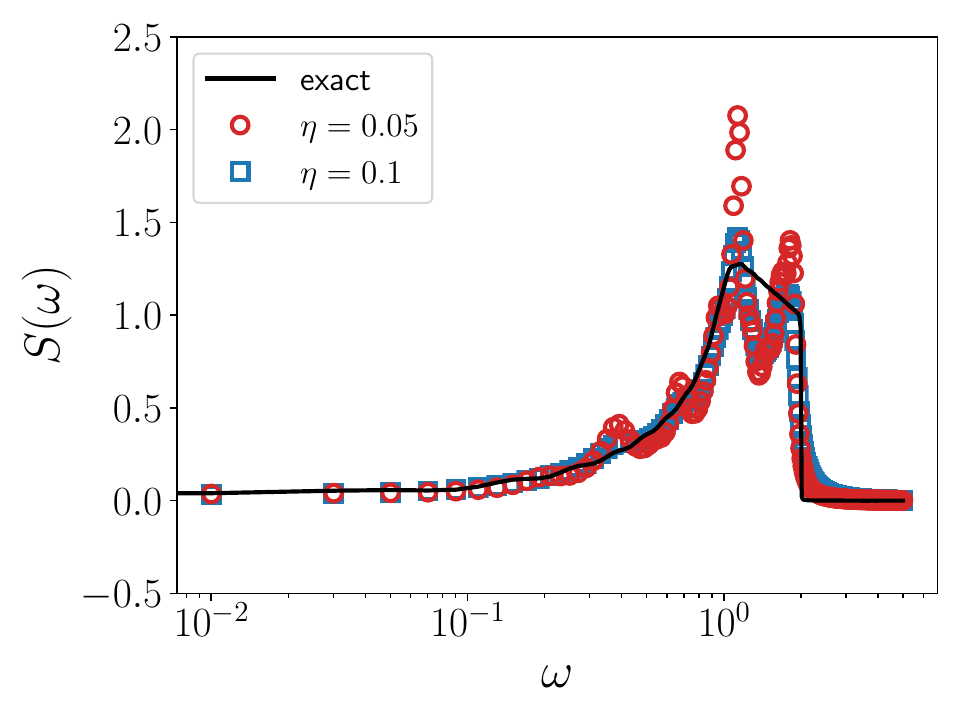}}
  \caption{The local spectral function of the XY chain computed in the same settings as Fig.~\ref{fig:xxz}(d) of the main texts. 
  \label{fig:Sw_logw}}
\end{figure}

Next, one obtains the dynamic susceptibility by performing analytic continuation. Substitute $i\omega \rightarrow \omega + i0^{+}$ in \Eq{eq:chi_matsubara} and taking the imaginary part, one has
\begin{equation}
\chi{''}(\omega) = -\frac{\pi}{Z}\sum_{m,n}  {\tilde{S}^z}_{nm} {\tilde{S}^z}_{mn} \left({e^ {-\beta \Lambda_m} - e^ {-\beta \Lambda_n} }\right)  \delta\left( \omega - \Lambda_m + \Lambda_n  \right). 
\end{equation}
In practical calculations we replace the delta function by a Lorentzian with broadening $\delta(x) = \lim_{\eta \rightarrow 0}\frac{1}{\pi} \frac{\eta }{x^2 + \eta^2}$. Figure~\ref{fig:Sw_logw} shows the local spectral function of the XY chain, in the same settings as Fig.~\ref{fig:xxz}(d) of the main texts. One sees that larger broadening factor $\eta$ smears out the spectral function. However, the low frequency spectrum is less sensitive to the broadening factor than the high frequency region. For the results in the main texts, we use $\eta = 0.05$. 

An alternative approach to perform analytic continuation is to numerically invert the following integral equation based on imaginary-time data:
\begin{equation}
\chi(\tau) = \int_{-\infty}^{\infty} \frac{d\omega}{\pi}  \frac{ \chi{''}(\omega)}{1-e^{-\beta \omega}} e^{-\omega \tau}. 
\end{equation} 
This is the starting point of the maximum entropy method and related approaches~\cite{Jarrell1996, Sandvik1998}. 

Finally, the dynamic susceptibility is related to the real frequency local spectral function 
via the fluctuation-dissipation theorem  $ S(\omega) = \frac{2 \chi{''}(\omega)}{1-e^{-\beta \omega}}$~\cite{SachdevQPT}. Since $\chi{''}(\omega)$ is an odd function of the frequency, one has $S(-\omega) = e^{-\beta \omega} S(\omega)$.  

\section{Klein bottle entropy} \label{app:klein}
The Klein bottle entropy is a quantity proposed in Ref.~\cite{Tu2017} to characterize
different CFTs in (1+1) dimensions.
It is defined by
\begin{equation}
S^{\mathcal{K}} = \ln \frac{Z^{\mathcal{K}} \left( 2 L, \frac{\beta}{2}
   \right)}{Z^{\mathcal{T}} (L, \beta)} ,
\end{equation}
where $L$ is the system size, $\beta$ is the inverse temperature, and $L \gg \beta \gg 1$. 
Here, $Z^{\mathcal{T}} =\tr (\mathrm{e}^{- \beta H})$ represents the partition function of a system with the periodic boundary condition.
The Klein bottle partition function $Z^{\mathcal{K}}$ is defined by inserting an extra operator $\Omega$ into $Z^{\mathcal{T}}$: $Z^{\mathcal{K}} =\tr (\Omega \mathrm{e}^{- \beta H})$. 
The operator $\Omega$ interchanges the left and right movers in the CFT Hilbert space, which, in lattice systems, is usually chosen as the bond-centered lattice reflection operator $P$~\cite{Tu2017,Tang2017,chen2017conformal}: 
\begin{equation}
P|s_1, s_2, \ldots, s_{L - 1}, s_L \rangle = |s_L, s_{L - 1}, \ldots, s_2,
   s_1 \rangle .
\end{equation}

\begin{figure}[!htb]
  \resizebox{\columnwidth}{!}{\includegraphics{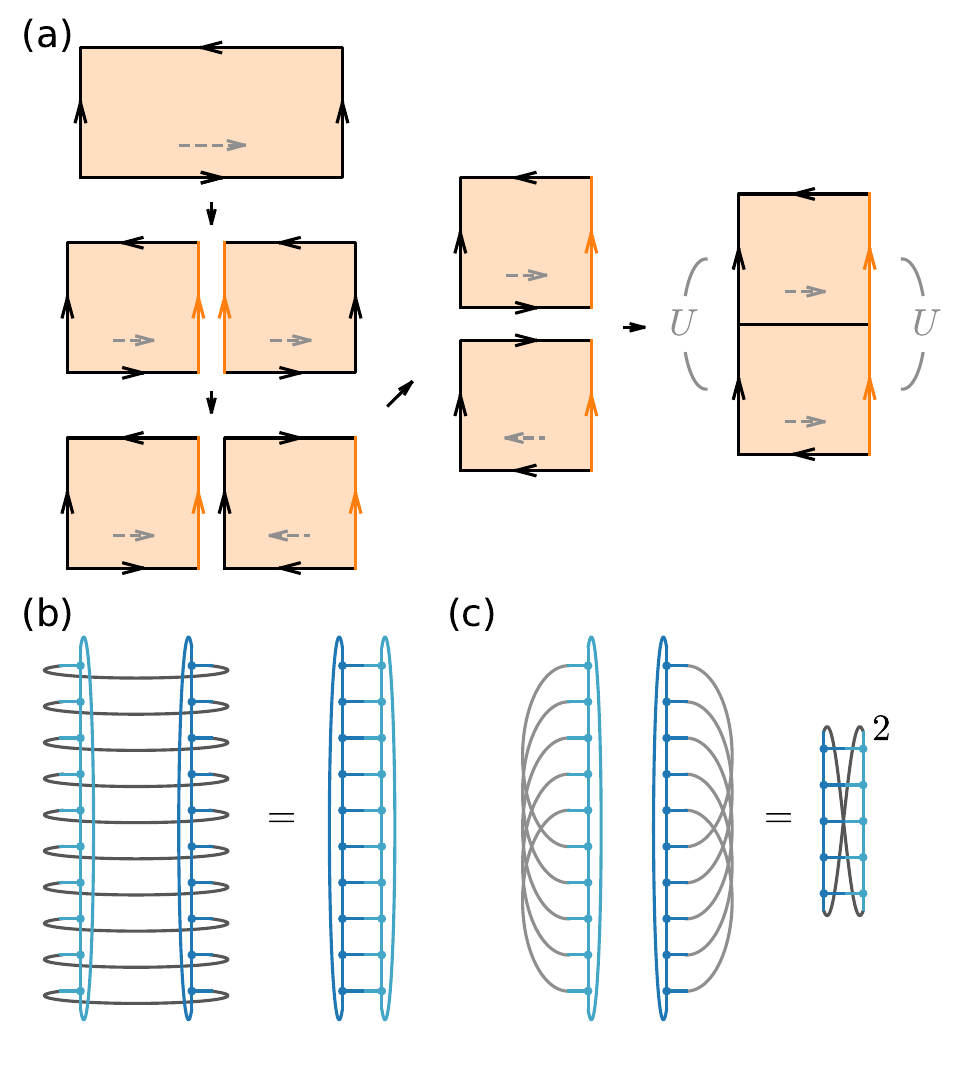}}
  \caption{(a) The cutting and sewing of the Klein bottle path-integral manifold. The dashed arrow indicates the ``direction'' of the spatial transfer matrix.
  (b) Calculation of the boundary entropy contained in the torus partition function $Z^{\mathcal{T}}$ using the boundary cMPS.
  (c) Calculation of the boundary entropy contained Klein bottle partition function $Z^{\mathcal{K}}$ using the boundary cMPS. Note that, here, each long-range interaction line contains a unitary gate $U$ that satisfies $(U\r)_i = \l_i$ and $(\l U)_i = \r_i$.
  \label{fig:klein-illustr}}
\end{figure}

Due to the existence of the reflection operator, the corresponding path integral manifold of $Z^{\mathcal{K}}$ is a Klein bottle, which is nonorientable, and thus is distinct from the torus (or cylinder, if the system has open boundary condition) corresponding to the ordinary partition function.
As shown in Fig.~\ref{fig:klein-illustr}(a), we cut the path integral into two halves, flip the right half spatially, and connect it to the left half from below. 
After these operations, the Klein bottle path integral manifold corresponding to spatial length $2L$ and inverse temperature $\beta/2$ becomes a cylinder that corresponds to spatial length $L$ and inverse temperature $\beta$. 
In this regard, the difference between $Z^{\mathcal{K}} \left( 2 L, \beta/2 \right)$ and $Z^{\mathcal{T}} (L, \beta)$ only lies at the boundaries of the system~\cite{Tang2017}. 
In this regard, the Klein bottle entropy can be taken as a boundary entropy, which can be calculated using the boundary cMPS in the cMPO formalism. 

A caveat of the calculation is that, during the cutting and sewing process, half of the path-integral manifold is flipped along the spatial direction [see Fig.~\ref{fig:klein-illustr}(a)].
When the cMPO is not Hermitian, the last step in Fig.~\ref{fig:klein-illustr}(a)---the reconnection of the two halves of path-integral manifolds---cannot be performed.
Hereby, we restrict our discussion to the cases where $\+$ can be connected to its conjugate transpose by a unitary transformation, i.e. $(\+_{ji})^* = (U \+ U^\dagger)_{ij}$.

One can verify that $U=U^\dagger$, and the local tensor $\r$ in the cMPS $|r\>$ is connected to the local tensor $\l$ in $\<l|$ by this unitary gate, i.e., $(U\r)_i = \l_i$ [c.f. Eq.~(\ref{eq:cmps}) of the main text].
Using this unitary transformation, we can flip the direction of $\+$ in one of the two halves and finish the reconnection in Fig.~\ref{fig:klein-illustr}(a).
As a result, each of the long-range interaction lines at the boundaries of the cylinder would contain a unitary gate $U$, as shown in the last step in Fig.~\ref{fig:klein-illustr}(a).

As mentioned above, the difference between $Z^{\mathcal{T}}$ and $Z^{\mathcal{K}}$ only locates at the boundaries, and the bulk contributions cancel with each other.
Therefore, we only need to consider the different contraction schemes of the boundary cMPS in the two cases.
Fig.~\ref{fig:klein-illustr}(b) shows the calculation of the boundary entropy for the periodic boundary condition, which is simply the overlap between $\<l|$ and $|r\>$.
To calculate the boundary entropy contained in $Z^{\mathcal{K}}$, we contract the boundary cMPS $\<l|$ and $|r\>$ separately, as shown in Fig.~\ref{fig:klein-illustr}(c).
Notice that each of the boundary long-range interaction lines contains a unitary gate $U$ that satisfies $(U\r)_i = \l_i$ and $(\l U)_i = \r_i$, and the Klein bottle entropy is calculated as
\begin{equation}
S^{\mathcal{K}} = 2 \ln \tr (\tilde{\Omega}
   \mathrm{e}^{-\beta K_{\gong} / 2}) - \ln
   \tr (\mathrm{e}^{-\beta
   K_{\gong}}),
\end{equation}
where $K_{\gong}$ is calculated with the local tensors $\l$ and $\r$ from $\<l|$ and $|r\>$ using Eq.~(\ref{eq:gong-mat}).
Here $\tilde{\Omega}$ is a swap operator that interchanges the indices corresponding to $\l$ and $\r$ [see Fig.~\ref{fig:klein-illustr}(c)]:
\begin{equation}
 \tilde{\Omega} = \left( \begin{array}{cc}
     & \id\\
     \id & 
   \end{array} \right), 
\end{equation}
where $\id=\id_{\l}=\id_{\r}$. 

\begin{figure}[!htb]
  \vspace{0.25cm}
  \resizebox{\columnwidth}{!}{\includegraphics{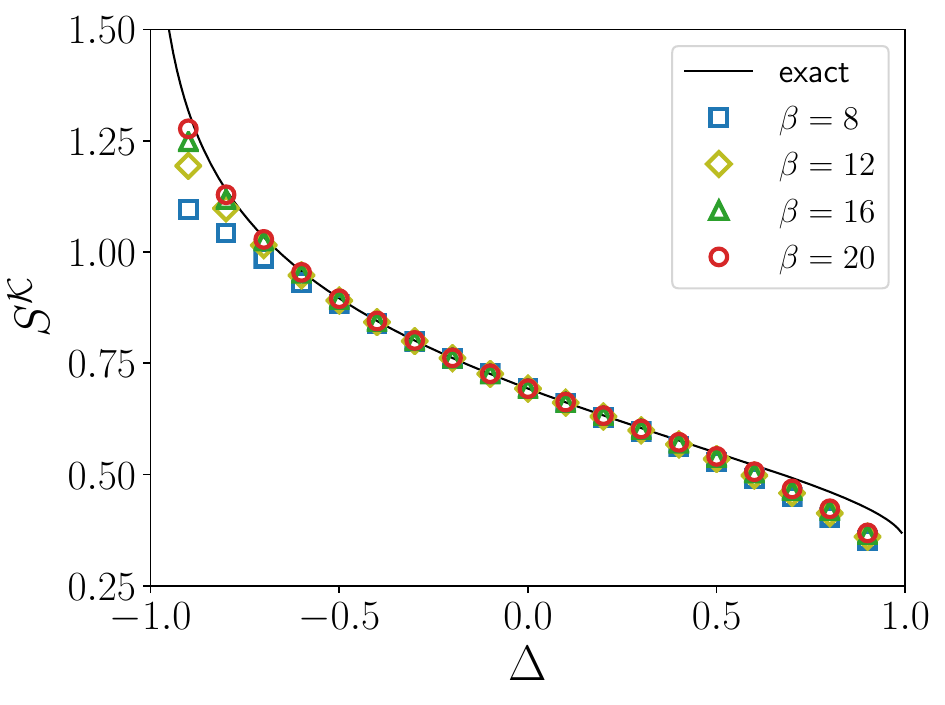}}
  \caption{The numerical results of the Klein bottle entropy in the critical phase of XXZ chain. The bond dimension of the boundary cMPS is $\chi=20$.
  \label{fig:klein-result}}
\end{figure}

As a demonstration, we calculate the Klein bottle entropy in the critical phase of the XXZ chain, whose Hamiltonian is given by Eq.~\eqref{eq:xxz-chain}.
For this model, in order to make the calculations in the lattice model consistent with the definition of Klein bottle entropy in CFT, we need to perform a unitary rotation to the basis such that the XY part of Heisenberg coupling becomes ferromagnetic.
In this critical phase where $-1 < \Delta \leq 1$, the XXZ chain is described by Luttinger liquid theory,
and the Klein bottle entropy $S^{\mathcal{K}}$ in this model is related to its Luttinger parameter $K$, which can be determined from the Bethe ansatz solution~\cite{tang2019klein}
\begin{equation}
 S^{\mathcal{K}} = \ln (2\sqrt{K}) = \ln (\sqrt{\frac{2\pi}{\pi-\cos^{-1}\Delta}}).
\end{equation}

Before the calculation, we perform a gauge transformation to the cMPO, 
so that it can be connected to its conjugate transpose by a unitary $U$:
\begin{equation}
 \+ = \left(\begin{array}{c|ccc}
     \id & \sqrt{ \dtau/2} S^+ & \sqrt{\dtau/2}  S^- &  \sqrt{\dtau \left|\Delta\right|} S^z\\ \hline
     \sqrt{\dtau/2} S^- &  &  & \\  
     \sqrt{\dtau/2} S^+ &  &  & \\
    - \mathrm{sgn}(\Delta) \sqrt{\dtau \left|\Delta\right|} S^z &  &  & 
     \end{array}\right) , \label{eq:xxz-cmpo-sym} 
\end{equation}
and the corresponding $U$ is given by
\begin{equation}
 U = \left(\begin{array}{c|ccc}
     1 &   &   & \\ \hline
       &   &  1& \\  
       &  1&   & \\
       &   &   & -\mathrm{sgn}(\Delta)
     \end{array}\right), \label{eq:xxz-cmpo-U} 
\end{equation}
where $\mathrm{sgn}(\Delta)$ represents the sign of $\Delta$.

As shown in Fig.~\ref{fig:klein-result}, the cMPO results are in good agreement with the exact solutions, except near $\Delta=-1$ and $\Delta=1$.
Near $\Delta=-1$, the numerical results show a large deviation when the temperature is not sufficiently low, but they show a clear tendency that they will approach the exact solution as $\beta$ becomes larger.
In the vicinity of $\Delta=1$, the slight deviations in the numerical results originate from the marginally irrelevant term at $\Delta=1$, which are also observed in QMC simulations~\cite{tang2019klein}.

\section{Exact results of the XY chain} \label{appendix:xy}

The XY model can be exactly solved by mapping to free fermions. 
The specific heat shown in Fig.~\ref{fig:xxz}(a) is for an infinite chain 
\begin{equation}
c = \frac{\beta^2}{2\pi}\int_{-\pi}^{\pi}  d q \frac{\cos^2 q \, e^{-\beta \cos q}}{(1+e^{-\beta \cos q})^2}.
\end{equation} 
For local dynamical properties shown in Fig.~\ref{fig:xxz}(b-d) we make use of the following analytical result for the time-dependent correlation function:
\begin{equation}
\langle S_i^z (t) S_i^z \rangle = \int_{- \pi}^{\pi} \frac{dkdq}{(2 \pi)^2} 
\frac{e^{i t (\cos k - \cos q)} e^{\beta \cos q}}{(1 + e^{\beta \cos k})  (1 +
e^{\beta \cos q})},
\end{equation} 
from which one can derive the imaginary-time spin correlator $\chi(\tau)= \<S^z_i (-i\tau) S^z_i \> $, the local susceptibility $\chi_{\mathrm{loc}}=\int_0^\beta d\tau \chi(\tau)$, and the local spectral function $S(\omega) = \int_{-\infty} ^{\infty} dt\<S^z_i (t) S^z_i \> e^{i\omega t}$. 

\section{Benchmark in the critical quantum Ising chain} \label{appendix:compare}

Here, in the critical quantum Ising chain, we benchmark the cMPO approach with the recently proposed differentiable tensor renormalization group ($\partial$TRG) algorithm~\cite{Chen2019m}.
When applied to quantum systems at finite temperature, the \dTRG approach starts from the MPO representation of the density matrix operator at a very high temperature and then lowers the temperature exponentially by repeatly doubling the density matrix operator on itself.
After each doubling step, the MPO is truncated by inserting isometries if its bond dimension exceeds the target bond dimension.
Using the automatic differentiation technique, \dTRG performs a global variational optimization on these isometries.
 
Depending on the depth $n_d$ of layers of isometries that is optimized during the temperature-lowering process, \dTRG shows different performances. 
For the cheapest choice $n_d=1$, \dTRG already outperforms the conventional approach, the linearized tensor renormalization group (LTRG), and its performance improves as $n_d$ increases~\cite{Chen2019m}.
In our benchmark, we will compare the performance of cMPO with \dTRG for both $n_d=1$ and $n_d=4$. 
The \dTRG code used in the benchmark is obtained from \footnote{See \url{https://github.com/TensorBFS/dTRG}}.

In the critical Ising chain, we compute the relative error of the free energy for both cMPO and \dTRG for two temperatures $\beta=32$ and $\beta=128$.
The \dTRG calculation is initialized at temperature $\beta=2^{-16}$ with a second-order Trotter decomposition, thus it has a Trotter error of the order $10^{-10}$, which is negligible.
The bond dimension in \dTRG corresponds to the cut-off dimension of the MPO
, while the bond dimension of the cMPO approach corresponds to the bond dimension of the periodic boundary cMPS, which corresponds to squared representability. 
In this regard, the bond dimensions of cMPO range from 4 to 28, and, correspondingly, we use bond dimensions from 16 to 784 for \dTRG (for $n_d=4$ we only calculate up to the bond dimension 400).

\begin{figure}[!htb]
  \resizebox{\columnwidth}{!}{\includegraphics{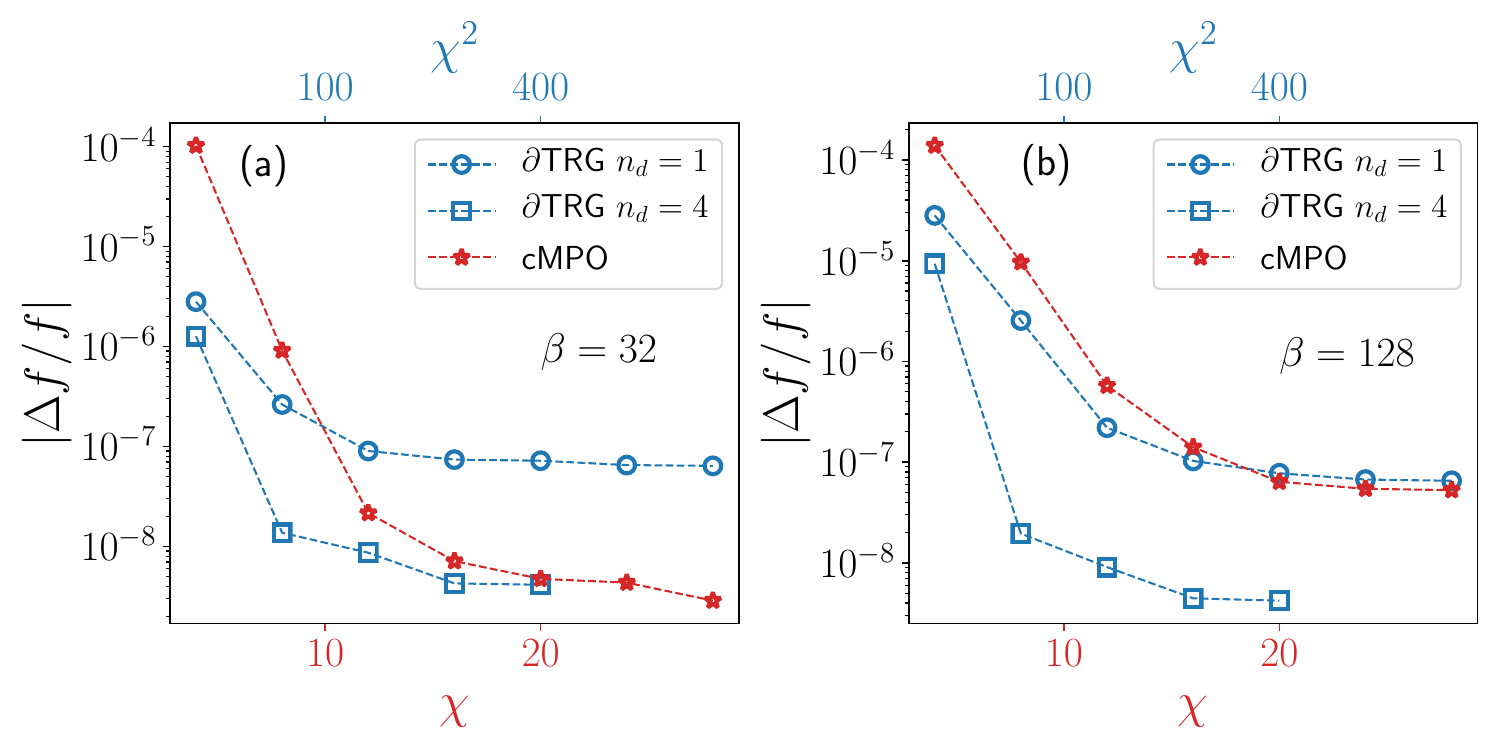}}
  \caption{Benchmark between cMPO and \dTRG in critical Ising chain. The parameter $n_d$ controls the depth of the isometries that are optimized~\cite{Chen2019m}. 
 The results are shown for (a) $\beta=32$ and (b) $\beta=128$.
  The horizontal axis at the bottom represents the bond dimension $\chi$ of cMPO approach, and, correspondingly, the horizontal axis at the top represents the bond dimension $\chi^2$ of \dTRG.
  }
  \label{fig:benchmark-ising}
\end{figure}

In Fig.~\ref{fig:benchmark-ising}, we show results for cMPO approach with bond dimension $\chi$ and \dTRG with bond dimension $\chi^2$.
At $\beta=32$, the cMPO approach apparently outperforms \dTRG with $n_d=1$ and is comparable with \dTRG with $n_d=4$ around bond dimension $\chi=20$, although it gives poorer results than \dTRG for small bond dimensions. 
At the lower temperature $\beta=128$, the accuracy of the cMPO approach is only comparable with \dTRG with $n_d=1$, and is visibly worse than \dTRG with $n_d=4$. 
We attribute this to the poor choice of the initial guess for boundary cMPS at this low temperature.
In our calculation, as discussed in Appendix \ref{appendix:initialstate}, for the smallest bond dimension $\chi=4$, we use boundary sites to form an initial boundary cMPS as the starting point. For other larger bond dimensions, we initialize the calculation with the optimized boundary cMPS with smaller bond dimensions. 
This initialization strategy is more suitable at high temperatures due to the short correlation lengths.
While at low temperatures, it is not sufficiently good, which makes the optimization trapped in some local minimum and leads to poorer results.
\end{document}